# Interlayer interactions reshape charge-density wave through electronic elasticity in $4H_b$-$TaS_2$

Carolina A. Marques[1,*], Martha Rösler[1], Berk Zengin[1], Aleš Cahlík[1], Edoardo Martino[2], Konstantin Semeniuk[2,3], Helmuth Berger[2], László Forró[4], Ana Akrap[5] and Fabian D. Natterer[1,*]

[1]Department of Physics, University of Zurich, Winterthurerstrasse 190, CH-8057 Zurich, Switzerland
[2]École Polytechnique Fédérale de Lausanne (EPFL), Institute of Physics, CH-1015 Lausanne, Switzerland
[3]Institute for Quantum Materials and Technologies, Karlsruhe Institute of Technology, Karlsruhe, Germany
[4]Department of Physics and Astronomy, University of Notre Dame, 225 Nieuwland Science Hall, IN 46556, USA
[5]Department of Physics, University of Zagreb, Bijenička cesta 32, 10000 Zagreb, Croatia

*correspondence shall be addressed to: cdamarques@outlook.com and fabian.natterer@physik.uzh.ch

## Abstract

Incommensurate charge-density waves (CDWs) in layered quantum materials frequently exhibit widely varying ordering wave vectors, even among nominally identical samples, obscuring their intrinsic electronic properties. Here we identify an inherent origin of this variability through the electronic elasticity of an incommensurate CDW using the model heterostructure $4H_b$-$TaS_2$, composed of alternating commensurate CDW on 1T and incommensurate CDW on 1H layers. Low-temperature scanning tunneling microscopy, combined with Fourier and quasiparticle-interference analysis, exploits the lattice-pinned 1T CDW as an internal reference to resolve discrete compressive (−2.3%) and tensile (+3.2%) elastic states of the neighboring 1H CDW selected by the interlayer registry of the adjacent layers. Corresponding few-meV shifts of a flat band demonstrate that weak interlayer interactions reshape the low-energy electronic structure through the intrinsic elasticity of the incommensurate CDW. These findings establish electronic elasticity as a mechanism by which subtle interlayer interactions control correlated electronic states in van der Waals heterostructures.

## Introduction

Charge density waves (CDW) in low-dimensional quantum materials are characterized by an ordering wave vector, accompanied by periodic modulations of the electronic density[1,2]. While commensurate CDW are locked to the lattice, incommensurate CDW are more flexible[3,1]. A persistent challenge is the report of widely varying incommensurate CDW wave vectors[4] in nominally identical samples, obscuring the relationship between intrinsic physical properties and the role of layer interactions.

This variability is commonly attributed to disorder or strain[5–10]. However, systematic dependencies of CDW wave vectors on layer thickness, substrate, and local environment in van der Waals materials have been reported, including $TaS_2$[11] and for $NbSe_2$ based systems[11,12]. This suggest that the incommensurate CDW wave vector itself is a soft degree of freedom adapting to more than just disorder, including residual strain and sample history[13–15].

Here we use the layered heterostructure $4H_b$-$TaS_2$[10,16] as a model system to investigate the variability of an incommensurate CDW. $4H_b$-$TaS_2$ hosts multiple coexisting quantum phases including superconductivity, commensurate and incommensurate CDW orders that are modified by interlayer interactions. Figure 1(a) shows the stacking sequence of 1T and 1H layers that characterizes the parent compound $4H_b$-$TaS_2$. The 1T layers stabilize the well-known “Star-of-David" CDW, which below 315 K[17] is locked to the $TaS_2$ lattice through its commensurate $\sqrt{13} \times \sqrt{13}$ phase and known for Mott physics in the pure 1T-$TaS_2$ allotrope[18–20]. Conversely, the H layers host a superconducting phase with the critical temperature of 0.7 K and an incommensurate CDW in bulk 2H-$TaS_2$[21,22]. When stacked between 1T layers, as a consequence of interlayer charge transfer from the 1T to the 1H layers,[23–28] the superconducting critical temperature in $4H_b$-$TaS_2$ becomes 2.7 K[29], and increases up to 5 K under hydrostatic pressure[15,30]. The incommensurate CDW in the 1H layers appears below 22 K, lower than in 2H-$TaS_2$,[16,31] with a mean wave vector close to 2.8×2.8, often labelled as 3×3 (see Supplementary Table S1). Upon cleaving, either layer can terminate the surface, resulting in large area terraces of each polymorph[17,32–36].

In this work, we use the lattice-pinned commensurate $\sqrt{13} \times \sqrt{13}$ CDW of the 1T layers as a rigid electronic reference to probe the incommensurate CDW in adjacent 1H layers. We show that the ordering wave vector of the 1H CDW responds elastically to the stacking registry of the surrounding commensurate CDW layers, giving rise to discrete compressive and tensile states that span nearly the full range of wave vectors reported in transition-metal dichalcogenides. These electronic deformations are accompanied by few-meV shifts of an unoccupied flat band, coming from the weak overlap of the "Star-of-David" demonstrating that interlayer-controlled CDW elasticity directly reshapes the low-energy electronic structure.

Together, these results establish that the ordering wave vector of an incommensurate CDW is an intrinsic elastic degree of freedom that is controlled by subtle interlayer interactions. More broadly, our results demonstrate that weak changes in interlayer registry can reshape the electronic ground state of layered quantum materials, providing a general framework for understanding how stacking controls correlated electronic phases in van der Waals homo and heterostructures.

## Results

Large-scale topographic imaging of the 1T-terminated surface of $4H_b$-$TaS_2$ reveals regions of distinct electronic contrast. Figure 1(b) shows a representative field-of-view (FOV) in which a 1T-terminated surface exhibits the characteristic commensurate $\sqrt{13} \times \sqrt{13}$ CDW pattern [37–44]. Indicated by the black arrow along the diagonal there is an atomically smooth transition between two distinct regions that we call D1 and D2 (see also Supplementary Figure S1). While the D1 region has a seemingly uniform intensity CDW, the D2 region displays a qualitatively different motif with a large real-space unit cell, consisting of a bright central feature surrounded by a darker ring. The subtle topographic contrast of D2 is consistent with previous observations[44] and emphasized in spectroscopic maps, as shown further below. The deep-dark spots are sparsely distributed atomic defects locally disrupting the Star of David pattern in both regions. The defect density of 0.010 ± 0.001 defects/nm$^2$ is found for both regions, excluding them as the origin of the distinct contrast between D1 and D2.

As illustrated schematically in Figure 1(c), we will show that regions D1 and D2 correspond to distinct elastic responses of the subsurface 1H incommensurate CDW to the stacking of the CDW between its two adjacent 1T layers. The CDW on the 1T-layers can have two distinct orientations, referred to as $\alpha$ and $\beta$ rotation in previous reports[10], having the CDW unit cell rotated by either +13.9° or -13.9° in relation to the atomic Ta lattice. These two orientations are observed in electron diffraction images of $4H_b$-$TaS_2$ (see Supplementary Figure S2). It results in two different stackings, either the CDW on the 1T-layers enclosing the 1H-layer have the same orientation, or each 1T layer has the CDW rotated in opposite directions, resulting in an overall angle between the CDW unit cells of 27.8°. These two stackings, together with the incommensurate CDW, will result in distinct moiré patterns, that we detect in STM.

To determine whether the stacking of the surrounding commensurate 1T layers modifies the buried incommensurate CDW, we compare three representative configurations (Figure 2): a 1H-terminated surface, which provides the reference wave vector of the incommensurate CDW, and the two 1T-terminated regions D1 and D2 identified in Figure 1. As shown below, these three measurements establish that the buried 1H CDW adopts two distinct elastic states depending on the relative orientation of the adjacent commensurate CDWs.

To quantify the stacking of the 1T layers and the respective elastic deformations of the incommensurate CDW in the 1H layer in between, we analyse the Fourier transforms of local density of states (LDOS) maps and topographic images taken over large FOVs, having high resolution to detect long range moiré patterns. In Figure 2 we show a comparative analysis of three representative cases: (a) a 1H-terminated surface, and (b,c) 1T-terminated surfaces corresponding to regions D1 and D2, respectively. Each panel in Figure 2(a-c) is a cut-out of larger LDOS images (Supplementary Figure S3), at energies where the contrast is enhanced and the distinct real-space patterns due to the different periodicities of the CDWs and resulting moiré patterns are visible. The Fourier transforms of the large FOV images reveal multiple sets of Bragg peaks associated with the q-vectors of the commensurate 1T CDW ($\boldsymbol{q}_{T00}$, $\boldsymbol{q}_{0T0}$, $\boldsymbol{q}_{00T}$), incommensurate 1H CDW ($\boldsymbol{q}_{H00}$, $\boldsymbol{q}_{0H0}$), moiré between the CDWs ($\boldsymbol{q}_{HT0}$, $\boldsymbol{q}_{TH0}$, $\boldsymbol{q}_{0HT}$, $\boldsymbol{q}_{T0T}$), and super moiré combinations ($\boldsymbol{q}_{TH0\text{-}T0T}= \boldsymbol{q}_{TH0}- \boldsymbol{q}_{T0T}$), indexed as $\boldsymbol{q}_{ijk}$, where i,j,k=H,T indicate the polymorph of the top, middle and bottom layer of the three layer stack, respectively. Their assignment is in excellent agreement with moiré pattern simulations (see Methods and Supplementary Section S5).

We first establish the reference wave vector of the incommensurate CDW from a 1H-terminated surface. Figure 2(d), shows the previously reported rich pattern with multiple Bragg peaks[45,41,46]. It originates from q-vectors of the surface atomic lattice, the incommensurate 1H CDW ($\boldsymbol{q}_{H00}$, grey), the subsurface commensurate 1T CDW ($\boldsymbol{q}_{0T0}$, blue), from their moiré pattern ($\boldsymbol{q}_{HT0}$, black), and from higher order combinations (see Supplementary Fig.S5). We extract $q_{0T0}$ = 0.60 Å$^{-1}$ for the q-vector of the commensurate 1T CDW, as expected for the $\sqrt{13}\times\sqrt{13}$ unit cell, and $q_{H00}$= 0.78 Å$^{-1}$ for the q-vector of the incommensurate 1H CDW, rotated by 13.9° from $\boldsymbol{q}_{0T0}$, revealing a 2.77×2.77 CDW unit cell, consistent with earlier reports[40,41,43,44]. To confirm the assignment of the moiré peaks $\boldsymbol{q}_{HT0}$, we simulate the moiré pattern[47] using the extracted values of $\boldsymbol{q}_{0T0}$ and $\boldsymbol{q}_{H00}$ (full details in Supplementary Section S5), which shows good agreement with the experimental Fourier transform, Figure 2(d) right panel.

We can use the q-vector of the commensurate 1T CDW as a reference to characterize the q-vector of the underlying 1H layer when measuring on a 1T-terminated surface. Region D1 corresponds to a compressed incommensurate CDW [Figure 2(e)] which exhibits the simplest Bragg-peak structure, with a single set of peaks at $q_{T00}$ = 0.60 Å$^{-1}$ (blue circle) and at ½$q_{T00}$ = 0.30 Å$^{-1}$. Although this configuration resembles previously reported surface terminations[38–41,48], the observed 2×2 order can be accounted for by the moiré pattern ($q_{TH0}$ = ½$q_{T00}$, black square) between the 1T CDW and a 1H incommensurate CDW with a q-vector of magnitude $q_{0H0}$ = 0.80 Å$^{-1}$ (grey circle), and rotated by 19.2° from $\boldsymbol{q}_{T00}$, as revealed by a good agreement with simulations. Such a $\boldsymbol{q}_{0H0}$ corresponds to a CDW that is contracted in real space by (-2.3 ± 0.5)%, and rotated by 5.3°, in relation to the CDW observed at the 1H-surface, and the single set of peaks $\boldsymbol{q}_{T00}$ is consistent with the surrounding 1T layers being aligned, as illustrated in panel (h).

Region D2 instead exhibits a tensile CDW state. The complex Bragg-peak pattern in Figure 2(f), indicate additional contributions from interlayer interactions involving the 1T layer beneath the 1H subsurface layer. We now observe two sets of peaks at $q_{T00} = q_{00T} = 0.60$ Å$^{-1}$ rotated by 27.8º relative to each other, with their mutual combination creating one moiré q-vector ($q_{T0T} = 0.29$ Å$^{-1}$, yellow) and their individual combinations with the 1H layer generating two other moiré q-vectors ($q_{TH0} = q_{0HT} = 0.22$ Å$^{-1}$, black and purple). These two moiré patterns also create a moiré between them, resulting in a super moiré q-vector ($q_{TH0\text{-}T0T} = 0.084$ Å$^{-1}$, green). Our simulations reproduce this assignment for higher order Bragg peaks as well, allowing us to quantify the q-vector of the incommensurate 1H CDW as $q_{0H0} = 0.75$ Å$^{-1}$ (grey circle), being expanded by (+3.2 ± 0.5)% in relation to the CDW q-vector at the 1H-terminated surface, as sketched in panel (i).

Taking the q-vector of the incommensurate CDW on the 1H-surface [Figure 2(g)] as a reference, in region D1, the CDWs of the adjacent 1T layers are crystallographically aligned and the incommensurate 1H CDW is compressed [Figure 2(h)], while in region D2, the CDW of the surrounding 1T layers are rotated in opposite directions leading to a tensile expansion of the 1H incommensurate CDW [Figure 2(i)].

Having established that the incommensurate CDW q-vector on 1H layers changes due to an intrinsic elastic response to interlayer interactions with the surrounding 1T-layers, we investigate the consequences for the electronic structure at the 1T-terminated surfaces. In Figure 3 we compare quasiparticle-interference (QPI) maps acquired in regions with compressive (D1) and tensile (D2) incommensurate CDW deformation on the 1H layers. Panels (a–d) show QPI maps measured in the two regions at selected bias voltages. The overall QPI patterns remain similar in the two regions, indicating that the principal electronic structure is preserved despite the different elastic states. We observe chiral QPI features in Figure 3(d) as counterclockwise twisted intensity clouds emerging around the $\boldsymbol{q}_{T00}$, which are preserved between regions D1 and D2 and unaffected by the CDW elasticity.

The energy line cuts over a wide energy range across the QPI maps along the $\boldsymbol{q}_{T00}$ direction, Figure 3(e), show that both regions are metallic, consistent with recent angle-resolved photoemission spectroscopy measurements[28]. However, only minor differences in the energy dispersion of the QPI features between regions D1 and D2 are observed, mainly in the unoccupied states between 200 and 400 mV, as also reflected in the averaged LDOS spectra shown in panel (f). The most significant differences in the dispersion of the electronic states between the two regions D1 and D2 emerge at low energies. A close-up near the Fermi level, Figure 3(g), reveals flat dispersions at 14 mV and 5 mV in region D1, corresponding to two peaks in the averaged LDOS spectrum in Figure 3(h). In region D2, two flat dispersions at are seen at lower biases, at 4 mV and 0 mV, corresponding to a board peak in the averaged LDOS [Figure 3(h)]. We identify these dispersions as coming from flat bands that in region D1 appear at higher energies, whereas in region D2 are shifted towards the Fermi level.

## Discussion

Our measurements establish that the q-vector of the incommensurate CDW on the 1H layers in $4H_b$-$TaS_2$ varies systematically with the stacking order of the commensurate CDW on the surrounding 1T-layers, and thus depends on the interlayer registry. We consider a minimal elastic model to quantitatively connect it to the observed changes in the electronic structure. The incommensurate CDW wave vector represents the elastic electronic order parameter whose wave vector $q$ minimizes a free energy $F(q, \theta)$, here expressed relative to $\boldsymbol{q}_{H00}$ from the incommensurate CDW on the 1H surface. $F(q, \theta)$ depends both on intrinsic electronic interactions and on interlayer registry, accounted for by the force $\lambda(\theta)$. To lowest order, $F(q, \theta)$ can be expressed as:

$$F(q,\theta) = \frac{K}{2}(q - q_{H00})^2 + \lambda(\theta)(q - q_{H00}).$$

Minimizing $F(q, \theta)$ leads to a connection of experimentally observed CDW strain to interlayer force and stiffness $\lambda/K$:

$$\frac{q(\theta) - q_{H00}}{q_{H00}} = -\frac{\lambda(\theta)}{Kq_{H00}},$$

Consequently, we understand interlayer registry as an effective force that shifts the equilibrium CDW wave vector. Here, elasticity refers to the finite free-energy cost associated with changing the equilibrium ordering wave vector of the incommensurate CDW. Within this framework, the compressive and tensile states observed in regions D1 and D2 correspond to opposite signs of $\lambda(\theta)$, set by the relative alignment or rotation of the CDW on the surrounding 1T layers.

This phenomenological description does not by itself fix the absolute energy scale of the CDW elasticity. We take the experimentally observed shift of the flat band, $\Delta E \approx 5$ meV, as a proxy for the sensitivity of the electronic structure to changes in $q$. Combining this with the measured wave vector variation $\Delta q \approx 0.02$ Å$^{-1}$, we obtain an effective force $\lambda \approx \partial E/\partial q \approx 0.2$-$0.3$ eVÅ, which yields an estimated stiffness $K \sim 10$ eVÅ$^2$. This corresponds to an elastic energy cost $\Delta F \sim K(\Delta q)^2/2$ of the order of a few meV for the observed 2–3% deformation.

Density functional theory (DFT) calculations for 1T-$TaS_2$ show that a flat band emerges at the Fermi level when accounting for the commensurate $\sqrt{13} \times \sqrt{13}$ CDW and including spin-orbit coupling[49,50], which is postulated to be the one that forms the Mott-insulating state. This state has been shown to be highly susceptible to the stacking order of the CDW on adjacent 1T layers in pristine 1T-$TaS_2$[51,52]. In $4H_b$-$TaS_2$, the charge transfer from the 1T to the 1H layers [53] has been shown to deplete this flat band according to DFT calculations on 1T/1H-$TaS_2$ bilayers[25] and bulk $4H_b$-$TaS_2$[54]. Those computations show that the amount of charge transfer sensitively depends on the interlayer distance, consistent with optical spectroscopy measurements[55]. Our QPI measurements show a flat dispersion above the Fermi level, compatible with an unoccupied flat band at the 1T-terminated surface. Accordingly, region D2 transfers less charge to the H-layer than region D1 where the flat band lies 5 meV above the Fermi level. The stacking between the 1H and the surrounding 1T layers mediates the charge transfer as well as the elasticity of the incommensurate CDW of the 1H layer and should result in changes to the interlayer distance between the 1T and 1H layers. These observations show that interlayer CDW stacking order can not only tune the ordering wave vector of incommensurate CDWs but also directly reshapes the low-energy correlated electronic states, and is compatible with the reported strain tunability of the superconducting critical temperature[15,30].

The elasticity of the incommensurate CDW demonstrated here is not unique to $4H_b$-$TaS_2$. In the broader literature, the prototypical van der Waals CDW material 2H-$NbSe_2$ is well known to exhibit an incommensurate CDW wave vector that varies with layer thickness, pressure, and local strain[56–58]. Within our framework, these previously reported variations can be naturally understood because of the finite stiffness of an incommensurate CDW, which accommodates weak interlayer or interfacial interactions through elastic adjustments of its ordering wave vector. This perspective provides a unified interpretation of spatially varying CDW periodicities and accompanying low-energy electronic reconstructions reported for $NbSe_2$[59] and related dichalcogenides.

The present work establishes that the wave vector of an incommensurate CDW behaves as an intrinsic elastic degree of freedom that can be tuned by subtle interlayer registry. While demonstrated here for the incommensurate CDW in $4H_b$-$TaS_2$, this principle is expected to extend to other layered quantum materials in which electronic order possesses a finite elastic stiffness. In such systems, weak changes in stacking or interfacial registry can shift the equilibrium wave vector of the ordered state and thereby reshape the low-energy electronic structure. More generally, our results show that the elasticity of electronic order provides a previously unrecognized pathway by which weak interlayer registry can control the electronic structure of layered quantum materials, opening new opportunities for engineering correlated phases in van der Waals homo and heterostructures.

## Methods

**Sample growth:**

Single crystals of $4H_b$-$TaS_2$ were grown using a chemical vapor transport method with iodine serving as the transport agent, operating between 820°C in the hot zone and 700°C in the cold zone. For the preparation of the $4H_b$ phase, pre-reacted $TaS_2$ powder was used, which had been synthesized from high-purity tantalum and sulfur. This powder was sealed in an evacuated quartz ampoule at a pressure of $10^{-8}$ torr and heated at 950 °C for two weeks. The $4H_b$ phase forms when the metal-to-chalcogen ratio is close to the nominal stoichiometry.

**STM measurements:**

We performed the measurements using a commercial low-temperature scanning tunneling microscope (CreaTec Fischer), operating at 1.2 K, unless otherwise specified. Similar to other transition metal dichalcogenides[60], we cleave the samples at room temperature and under ultra-high vacuum conditions by knocking-off a CuBe rod glued to the surface of the single crystals with EpotekH20E, right before transferring the sample into the STM head. The STM tip is made of a mechanically cut Pt/Ir wire, and prepared on atomically flat surfaces of single crystals of Ag(111) and Au(111) by slight indentation and pulsing up to 10 V. The tip is held at a virtual ground, and the bias voltage was applied to the sample. The tunneling spectroscopy was measured by two methods. The first method is the conventional lock-in technique where the bias voltage is modulated at a single frequency to measure $dI/dV$, at a lock-in frequency of 819 Hz. The second method uses parallel spectroscopy, described in Zengin et al.[61], where the bias is modulated with an AC amplitude as large as the energy range of interest at a fixed frequency of 640 Hz. Therein, nonlinearities in the I(V) curve produce integer harmonics in the current that are simultaneously measured up to 62$^{nd}$ order (Intermodulation Products, MLA-3) from which the $I(V)$ and the $dI/dV$ data can be computed. The $I_{max}$ parameter corresponds to the maximal tunneling current during the large amplitude excitation, typically corresponding to currents at the voltage extrema. The time per spectrum is 500 ms. To obtain high reciprocal space resolution, we measure large FOV using a combination of parallel spectroscopy and sparse sampling[61,62]. The data shown in Figure 2 and 3 was measured by randomly selecting 20% of the pixels from a 1024 by 1024 grid, with lower probability at the borders of the FOV thereby creating a windowing function. To minimize the tip-induced drift at each row of the map, the tip followed the typical scan movement of a topography, and we added a 50 s wait time at the start of each row. The total map time was 21 h, 52 h and 54 h for the data in Figure 2(a),(b) and (c), respectively. We observe the same moiré, spectral features, and van Hove singularity shifts with conventional spectroscopy (see Supplementary Figure S6).

**Moiré Analysis:**

The q-resolution $\delta q \approx 2\pi/L$ is given by the finite size $L$ of the STM field of view (without accounting for windowing), providing a lower bound of $\delta q = 0.004$ Å$^{-1}$ on the uncertainties in extracted wavevectors and explaining the benefit of large FOV imaging. The associated angular resolution is $\delta\Theta \approx \delta q/q = 2\pi/(Lq)$. To extract the value $q_{0H0}$ or $q_{H00}$ of the strained incommensurate CDW, we compare experimental Bragg peaks to moiré simulations following Ref.[63]. We construct higher order combinations of the 1T reference CDW ($\boldsymbol{q}_{T00}$, $\boldsymbol{q}_{0T0}$, or $\boldsymbol{q}_{00T}$) with a variable 1H q-vector ($\boldsymbol{q}_{H00}$, $\boldsymbol{q}_{0H0}$). This generates moiré wavevectors between 1T and 1H incommensurate CDW ($\boldsymbol{q}_{HT0}$, $\boldsymbol{q}_{TH0}$, $\boldsymbol{q}_{0HT}$), as well as higher order "moiré of the moiré" via super moiré contributions ($\boldsymbol{q}_{TH0\text{-}T0T}$). Each real space lattice is constructed using $f_\alpha(\boldsymbol{q}_\alpha, \boldsymbol{r}) = \frac{1}{9} + \frac{8}{9}\prod_{i=1}^{3}\cos(\boldsymbol{q}_{\alpha,i}\boldsymbol{r})$. Higher-order peaks of order $n$ are generated through products and powers of the composing lattices, $f_{Moire}(\boldsymbol{r}) = \prod_\alpha f_\alpha^n(\boldsymbol{q}_\alpha, \boldsymbol{r})$. The error bars for the incommensurate CDW vectors $\boldsymbol{q}_{H00}$ and $\boldsymbol{q}_{0H0}$ are determined by systematically detuning both $\boldsymbol{q}$ and $\Theta$ around their optimal values until the Bragg peak intensity overlap between simulation and experiment shows a local minimum. This leads to the lower bound errors reported in the figure captions.

We correct for image distortions caused by creep and unequal lateral piezo calibration using a global affine transformation enforcing threefold symmetry imposed by the 1T reference q-vectors ($\boldsymbol{q}_{T00}$, $\boldsymbol{q}_{0T0}$, or $\boldsymbol{q}_{00T}$). The QPI data are enhanced similarly to increase signal to noise and respect the crystal symmetry.

**Acknowledgments**:

We thank Árpád Pásztor, Titus Neupert, and Mark H. Fischer for fruitful discussions. C.A.M, M.R., B.Z, A.C. and F.D.N. thank the Swiss National Science Foundation for financial support (200021-232187, 206021_213238, PP00P2_211014). C.A.M. acknowledges funding through a Foreign Students for the Swiss Government Excellence Scholarship (ESKAS No. 2023. 0017). E.M., K.S., H.B. thank the Swiss National Science Foundation through its SINERGIA network MPBH and grants No. 200021_175836 and PP00P2_170544.

**Author contribution**

C.A.M. and F.D.N. conceived the project; C.A.M. performed STM measurements with assistance from B.Z., A.C., and F.D.N.; C.A.M. analyzed the STM data with assistance from M.R.; H.B. synthetized the samples; E.M., K. S., L.F. and A.A. provided and characterized the samples with TED; C.A.M. prepared the figures. F.D.N. and C.A.M. wrote the manuscript with contributions from all authors.

Fig.1:

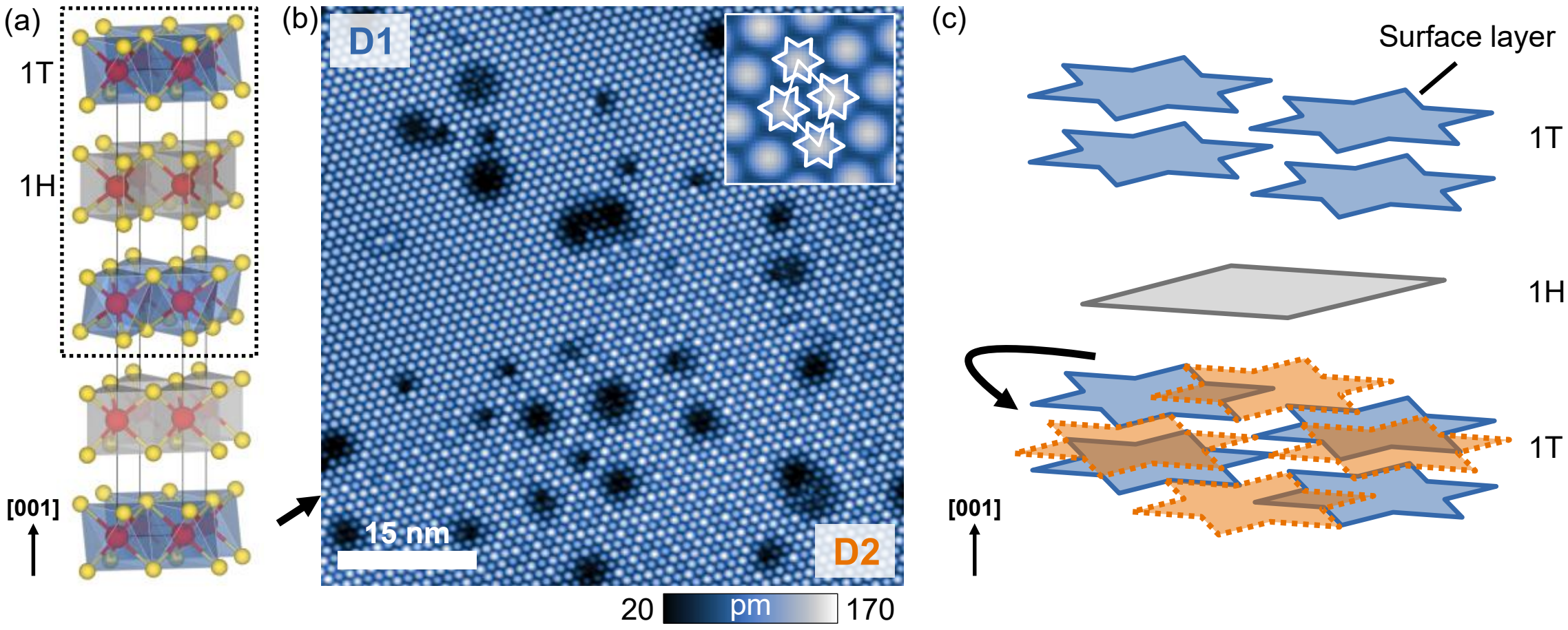

Figure 1 -**Real space super modulations at the 1T-surface of 4H$_b$-TaS$_2$.** (a) Unit cell of 4H$_b$-TaS$_2$, showing the alternating TaS$_2$ layers with 1T (blue) and 1H (grey) structural coordination. The dashed box indicates a three-layer stack consisting of two 1T layers and one 1H layer in between. (b) STM topography taken at the 1T-surface layer ($V_{set}$ = -50 mV, $I_{set}$ = 100 pA). The inset is a zoom-in on a 4x4nm$^2$ area showing the $\sqrt{13} \times \sqrt{13}$ charge density wave (CDW) of the 1T-layer, as indicated by the Star-of-David patterns. There are two regions with distinct super modulations superimposed over the CDW, D1 (upper left) and D2 (bottom right), with the boundary between them running along the direction indicated by the black arrow on the left. (c) Illustration of the CDW unit cells on a three-layer stack of 1T-1H-1T layers. The CDW on the 1T layers can be aligned (both blue) or rotated by 27.8° (blue and dashed orange).

Fig.2:

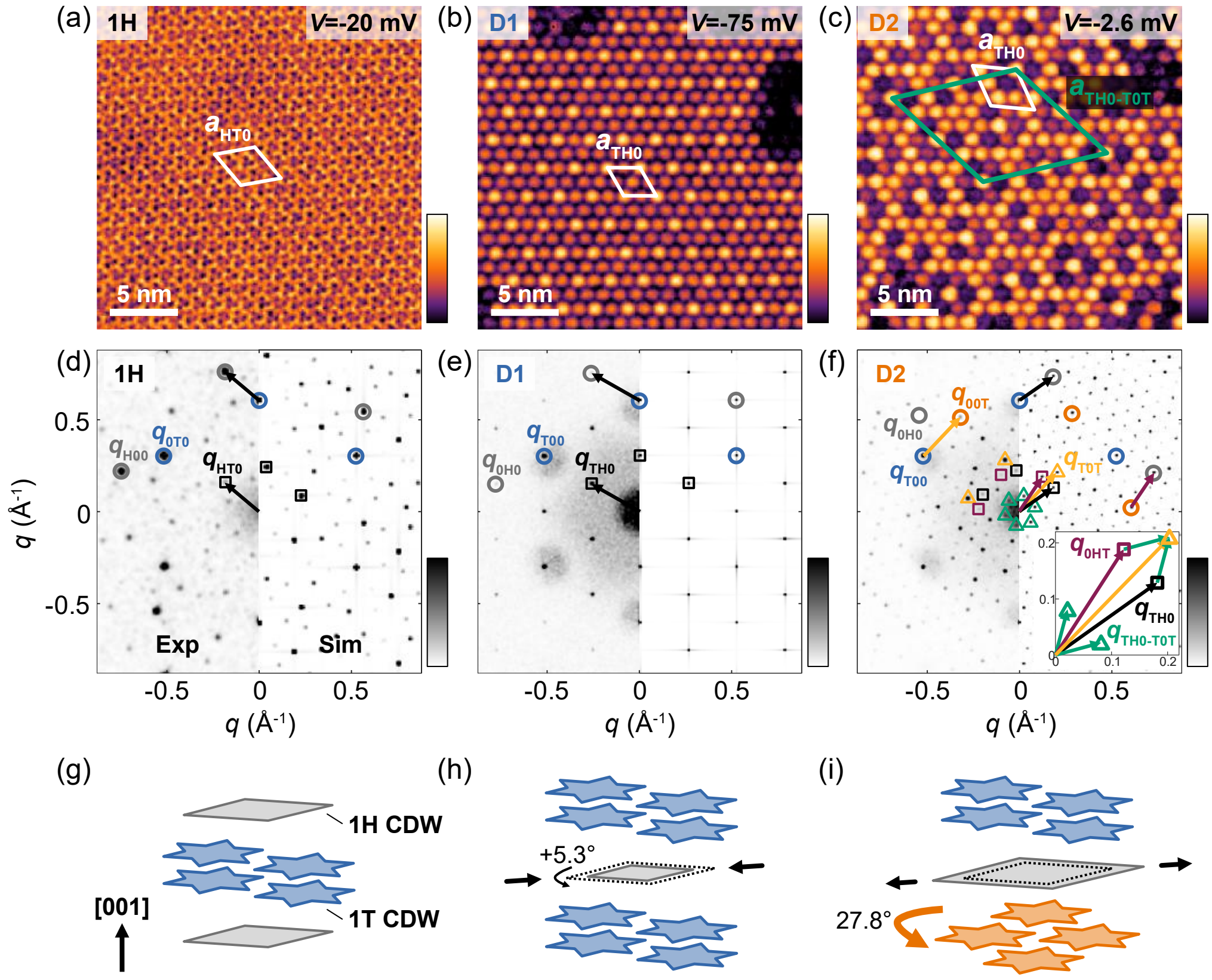

Figure 2 – **Quantifying elasticity of incommensurate charge density wave**. Differential conductance maps taken on $4H_b$-$TaS_2$ at (a) 1H-$TaS_2$ terminated surface, (b) 1T-$TaS_2$ terminated surface in region D1, and (c) in region D2. The images are rotated to align all 1T-CDW with the image frame and are cut-outs of larger areas (see full areas in supplementary section S4). The white rhombus indicates the unit cell of the moiré pattern, $a_{HT0}$ or $a_{TH0}$, between the incommensurate CDW on the 1H layers and the commensurate CDW on the 1T-layer. The large-area unit cell indicated by the green rhombus in (c) corresponds to the super moiré pattern, $\boldsymbol{a}_{TH0\text{-}T0T}$, between the moiré of the 1T-1H CDW, $\boldsymbol{a}_{TH0}$, and the moiré of the 1T-1T CDW of the surface and subsurface 1T layers, $\boldsymbol{a}_{T0T}$. (d-f) Fourier transform of the large area images corresponding to (a-c), respectively, after windowing and 6-fold symmetrization to increase signal-to-noise ratio. The right hand side of each panel (d-f) shows the moiré pattern simulation calculated as described in supplementary section S5. The characteristic q-vectors of the incommensurate CDW on the 1H layer (gray circle) and of the CDW on 1T layers (blue circle) are indicated on all panels. The black arrows and squares indicate the q-vectors of the moiré between the CDW of 1H and 1T layers ($\boldsymbol{q}_{HT0}$ or $\boldsymbol{q}_{TH0}$). The more complex pattern of (f) is due to multiple moiré structures and a super moiré. The former moiré are created by two rotational alignments of surface and subsurface 1T CDW with the middle 1H layer (black $\boldsymbol{q}_{TH0}$ and purple $\boldsymbol{q}_{0HT}$) and between themselves (yellow $\boldsymbol{q}_{T0T}$). The latter super moiré (moiré of moiré, $\boldsymbol{q}_{TH0\text{-}T0T}$) is shown in green. The inset in (f) is a close-up around $\boldsymbol{q}$=(0,0) to show how $\boldsymbol{q}_{TH0\text{-}T0T}$ is obtained as the moiré between $\boldsymbol{q}_{TH0}$ or $\boldsymbol{q}_{0HT}$ and $\boldsymbol{q}_{T0T}$. (g-i) Sketches of the relative stacking of the CDW on each 1T layer, their respective alignment, and CDW elasticity on the 1H layer. (Map parameters: (a,d) $f_{drv}$ = 640 Hz, $V_{drv}$ = 500 mV, $I_{max}$ = 685 pA, $\delta q$ = 0.009 Å$^{-1}$; (b,e) $f_{drv}$ = 640 Hz, $V_{drv}$ = 500 mV, $I_{max}$ = 1.2 nA, $\delta q$ = 0.004 Å$^{-1}$; (c,f) $f_{drv}$ = 640 Hz, $V_{drv}$ = 40 mV, $V_b$ = 35 mV, $I_{max}$ = 1.2 nA, $\delta q$ = 0.004 Å$^{-1}$).

Fig.3:

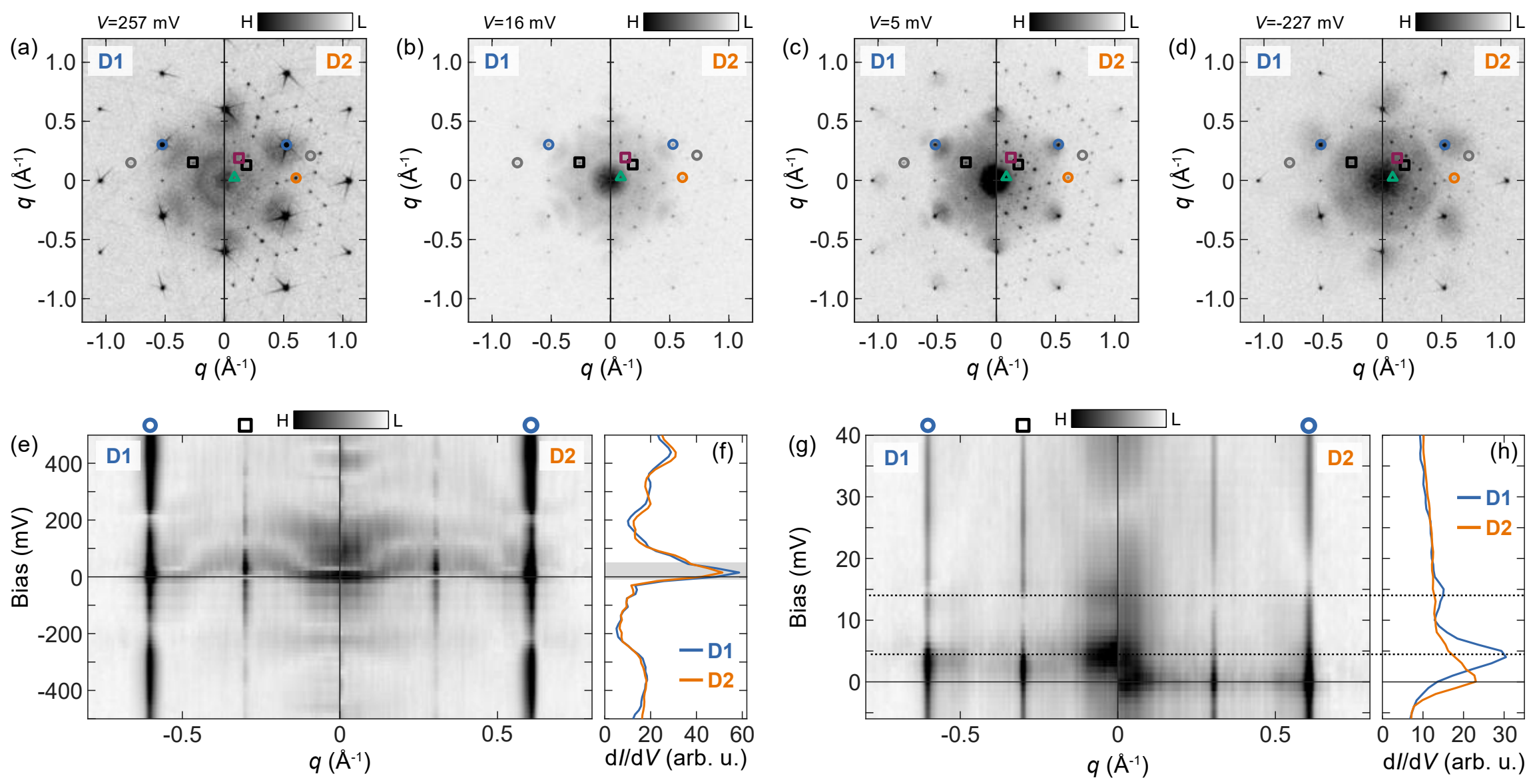

Figure 3 –**Consequences of incommensurate CDW strain on electronic structure**. (a-d) Fourier transform of differential conductance images taken at decreasing bias voltages $V$ = 257 mV, 16 mV, 5 mV and -227 mV, respectively, taken in regions D1 (left) and D2 (right). Despite distinct differences in the Bragg peak pattern, as explained in Figure 2, the QPI signatures between the two regions only show subtle differences. The circles and square symbols have the same meaning as in Figure 2. (e) Energy dispersion of the QPI signatures along $\boldsymbol{q}_y$ for $q_x$=0 Å$^{-1}$ taken in regions D1 (left) and D2 (right) over a wide energy range of ±500 mV. (f) Averaged differential conductance spectrum of the map on each region. (Map parameters: $f_{exc}$ = 640 Hz, $V_{drv}$ = 500 mV, $I_{max}$ = 685 pA). (g) Energy dispersion in a narrow energy range around the Fermi level corresponding to the shaded region in (f). While region D1 shows flat dispersions at 14 mV and 5 mV, in region D2 two flat dispersions are observed at 4 mV and 0 mV. This is also observed as a dispersion around the $\boldsymbol{q}_{T00}$ peaks (blue circles). (h) Averaged differential conductance spectrum of the map on each region, showing two peaks at 14mV and 5 mV for region D1, and a broader peak at 0 mV for region D2. (Map parameters: $f_{drv}$ = 640 Hz, $V_{drv}$ = 40 mV, $V_b$ = 35 mV, $I_{max}$ = 1.2 nA).

Supplementary material for '**Interlayer interactions reshape charge-density wave through electronic elasticity in $4H_b$-$TaS_2$**'

Carolina A. Marques[1,*], Martha Rösler[1], Berk Zengin,[1] Aleš Cahlík[1], Edoardo Martino[2], Konstantin Semeniuk[2,3], Helmuth Berger[2], László Forró[4], Ana Akrap[5], and Fabian D. Natterer[1,*]

[1]Department of Physics, University of Zurich, Winterthurerstrasse 190, CH-8057 Zurich, Switzerland
[2]École Polytechnique Fédérale de Lausanne (EPFL), Institute of Physics, CH-1015 Lausanne, Switzerland
[3]Institute for Quantum Materials and Technologies, Karlsruhe Institute of Technology, Karlsruhe, Germany
[4]Department of Physics and Astronomy, University of Notre Dame, 225 Nieuwland Science Hall, IN 46556, USA
[5]Department of Physics, University of Zagreb, Bijenička cesta 32, 10000 Zagreb, Croatia

## S1. Charge density wave q-vectors in TMDs

The H-type polymorphs of $NbSe_2$, $TaSe_2$ and $TaS_2$ have long been reported to develop a 3x3 CDW, that is $q_{at}/q_H$~3. However, these CDWs are rarely commensurate with the atomic lattice. Table S1 shows values of $q_{at}/q_H$ for different systems containing H-coordinated layers that develop such CDW. While the bulk 2H-systems show a slightly incommensurate CDW in $NbSe_2$ and $TaS_2$, in $TaSe_2$ the CDW becomes commensurate at low temperatures. When H-coordinated layers are intercalated with 1T-layers in the 4Hb structure, both in $4H_b$-$TaSe_2$ and $4H_b$-$TaS_2$ the CDW on the H layers is reported to be incommensurate with smaller value of $q_{at}/q_H$. It illustrates how this type of CDW is susceptible to the environment, and its's elasticity as we report in this work.

Table S1 - Compilation of values extracted from the literature of the periodicity of the CDW on H-coordinated layers in different polymorphs of $NbSe_2$, $TaSe_2$ and $TaS_2$.

| Material | $q_{at}/q_H$ | Ref. |
|---|---|---|
| 2H-$NbSe_2$ (bulk) | 3.04-3.06 | (*1–3*) |
| 2H-$TaSe_2$ | 3.06 (T>90K) | (*1*, *4*) |
| | 3.00 (T<90K) | |
| $4H_b$-$TaSe_2$ | 2.89 | (*5*) |
| 2H-$TaS_2$ | 2.95-2.97 | (*6*, *7*) |
| 1H/1T/2H-$TaS_2$ | 2.7+/-0.3 | (*8*) |
| $4H_b$-$TaS_2$ | 2.78-2.80 (H-surface) | This work and (*9*) |

## S2. Regions D1 and D2 with and without CDW domain boundary on the surface 1T layer

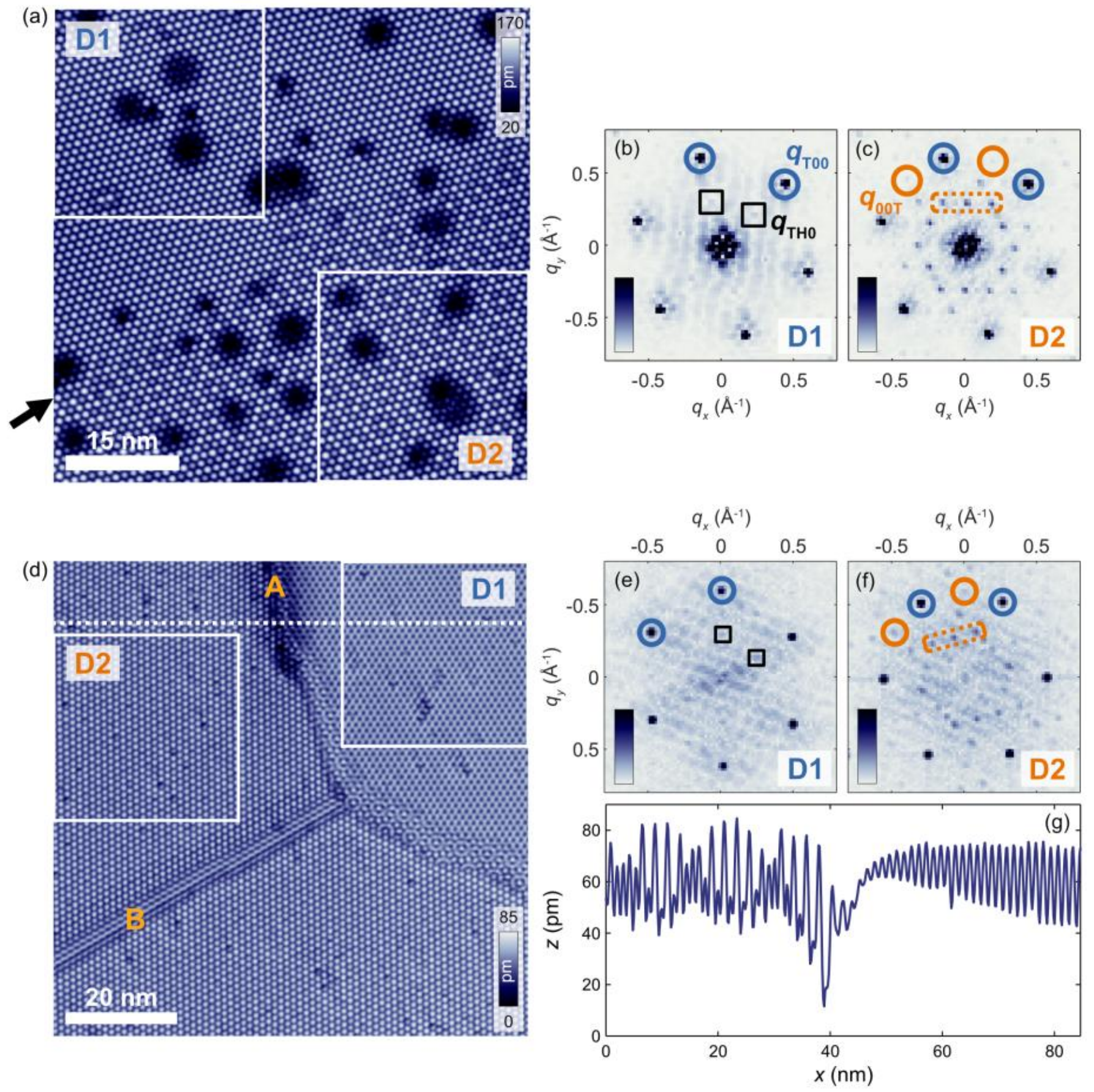

Fig. S1 – **Types of boundaries between regions D1 and D2.** (a) Topography of the 1T-terminated surface reproduced from Fig.1(b) of the main manuscript, showing regions D1 and D2 without a 1T-CDW domain boundary ($V_{set}$=-50 mV, $I_{set}$=100 pA). (b, c) Fourier transform of the areas D1 and D2 indicated by white squares in (a). The blue circles indicate the peaks corresponding to $\boldsymbol{q}_{T00}$ of the surface layer, orange circles to $\boldsymbol{q}_{00T}$ of the subsurface 1T layer, and black squares to $\boldsymbol{q}_{TH0}$,the moiré peak between 1T and 1H CDWs as defined in the main text. The dashed orange box in (c) indicates the characteristic pattern observed in region D2. (d) Topography of the 1T-terminated surface, showing two domain boundaries of the $\sqrt{13} \times \sqrt{13}$ CDW, A and B. Domain boundary A occurs due to rotation by 30° between the 1T-CDW on each side, while domain boundary B occurs due to a phase shift by one unit cell of the CDW ($V_{set}$=200 mV, $I_{set}$=100 pA). (e) and (f) show the Fourier transform taken over the areas D1 and D2 indicated as white squares in (d), at either side of domain boundary A. It shows that peak $\boldsymbol{q}_{T00}$ is rotated by 30° between the two images (blue circles). In addition to the change in rotation of $\boldsymbol{q}_{T00}$, (f) shows additional peaks compared to (e), the same as assigned in (c) and in Fig.2 of the main text. They are consistent with panels (b) and (c). (g) Line profile along the dashed line in (d), where no step edge is found, showing that the CDW domain boundary occurs in the same atomic layer.

In Fig.1(b) of the main manuscript, reproduced in Fig. S1(a), the transition between regions D1 and D2 occurs without a change to the $\sqrt{13} \times \sqrt{13}$ CDW of the surface 1T layer. This is shown by taking the Fourier transform over areas on each side of the boundary between the two regions, Fig. S1(b,c), where the peaks corresponding to $\boldsymbol{q}_{T00}$ are at the same positions on both regions D1 and D2. The differences between the Fourier transforms in Fig. S1(b,c)are the ones characterized in Fig.2 of the main manuscript corresponding to the characteristic peaks observed for regions D1 and D2, with the respective moiré patterns.

We found regions on other samples where the transition between regions D1 and D2 occurs across a domain boundary of the $\sqrt{13} \times \sqrt{13}$ CDW, indicated in Fig. S1(d) as domain boundary A.

Taking the Fourier transform over areas at each side of domain boundary A, Fig. S1(e,f), shows that the 1T-CDW is rotated by 27.8° between them, while the modulus of $\boldsymbol{q}_{T00}$ remains the same. Comparing Fig. S1(e,f) with Fig. S1(b,c), respectively, we can identify that the region on the left of domain boundary A corresponds to region D2 and on the right to region D1. In Fig. S1(f), the 27.8° rotated $\boldsymbol{q}_{T00}$ exposes a $\boldsymbol{q}_{00T}$ that is at the same position as $\boldsymbol{q}_{T00}$ in Fig. S1(e), showing that the subsurface 1T-layer is uniform over the area of the image, and that we can detect the CDW of that subsurface 1T-layer. A line profile across the CDW domain boundary A shows that this occurs on the same 1T-surface layer, as no sharp change in height is observed, Fig. S1(g), therefore not corresponding to a step edge.

A second domain boundary B is observed in Fig. S1(d), due to the phase shift by one unit cell of the 1T-CDW however does not result in a change in periodicity, with the characteristic peaks from region D2 identified on either side of boundary B.

The observation of the appearance of regions of type D1 and D2 over areas with and without a 1T-CDW domain wall on 1T surface layer, supports our interpretation described in the main manuscript: the differences in the Fourier peaks observed between region D1 and D2 come from the moiré pattern between aligned and misaligned CDWs on the 1T-$TaS_2$ layers sandwiching a 1H-layer. In this way, Fig. S1(a) shows the case where the top 1T-CDW is uniform, but there is a CDW-domain wall between two regions of 27.8° rotated 1T-CDW on the subsurface-layer, while Fig. S1(d) shows the case where the 1T-CDW of the subsurface 1T-layer has the same orientation over the area of the image, but there are CDW domains on the top 1T-surface layer.

### S3. Transmission electron diffraction image of 4Hb-$TaS_2$

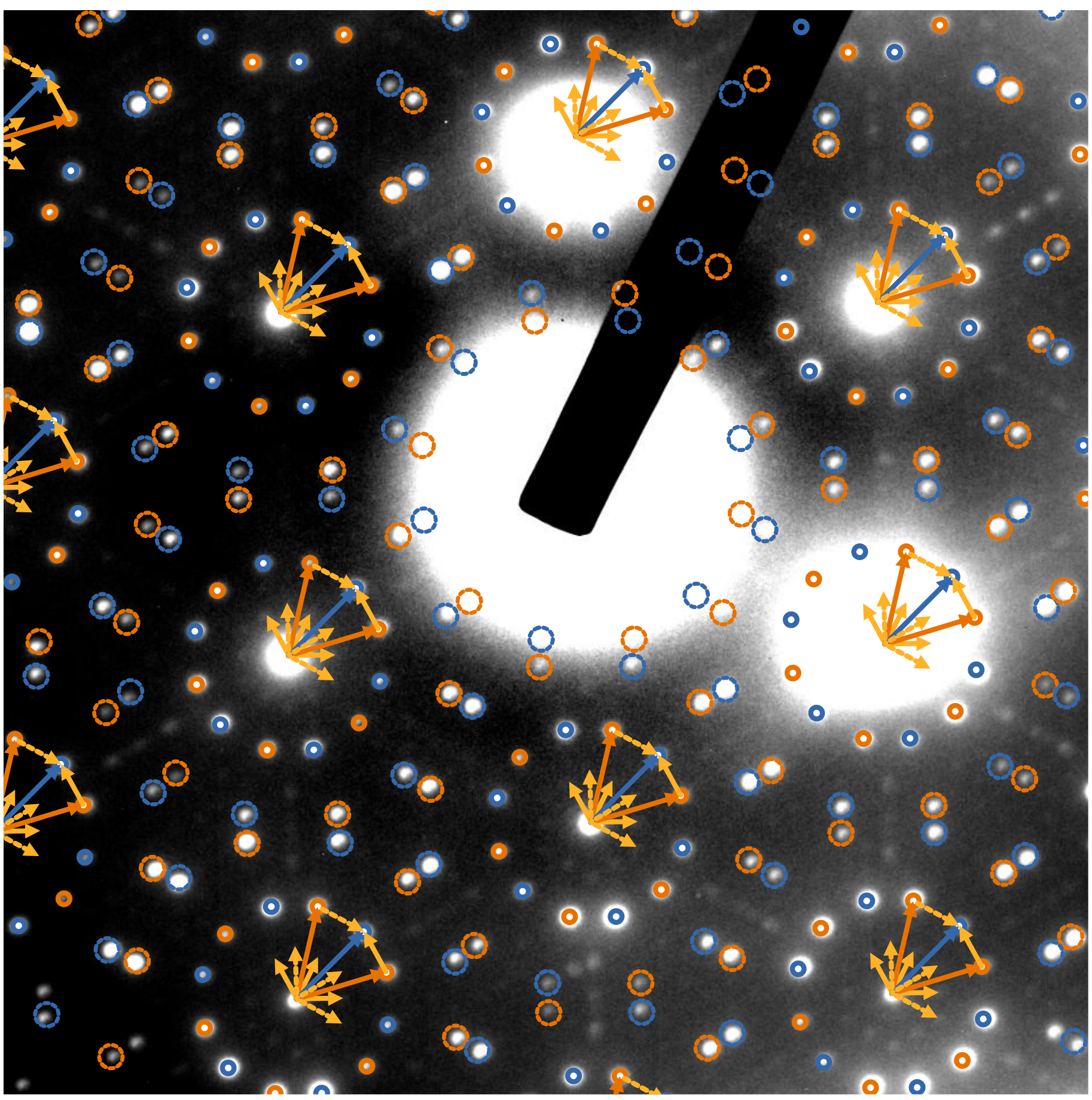

Fig. S2 – **Transmission electron diffraction image of bulk 4Hb-$TaS_2$.** Transmission electron diffraction image recorded at 4.8K. Two sets of satellite reflections are seen around the Bragg peaks of the atomic lattice, corresponding to two distinct orientations of the $\sqrt{13} \times \sqrt{13}$ CDW on 1T-layers, each orientation is indicated by blue and orange circles. The dotted blue and orange circles indicate higher order reflections of the same CDWs. The solid and dotted yellow arrows indicate the two possible moiré peaks, with the solid yellow arrow corresponding to $\boldsymbol{q}_{TOT}$.

Transmission electron diffraction (TED) images reveal satellite reflections consistent with two distinct orientations of the $\sqrt{13} \times \sqrt{13}$ CDW on the 1T-layers, blue and orange circles in Fig. S2. The angle between the two CDW orientations is 27.8°, consistent with the angle extracted from the STM measurements on region D2 (Fig.2(f) of main manuscript and Fig. S). This diffraction pattern shows that the CDW on the 1T-layers can form with two orientations rotated in respect to the atomic unit cell by ±13.9°, which are energetically equivalent.

## S4. Wider area images

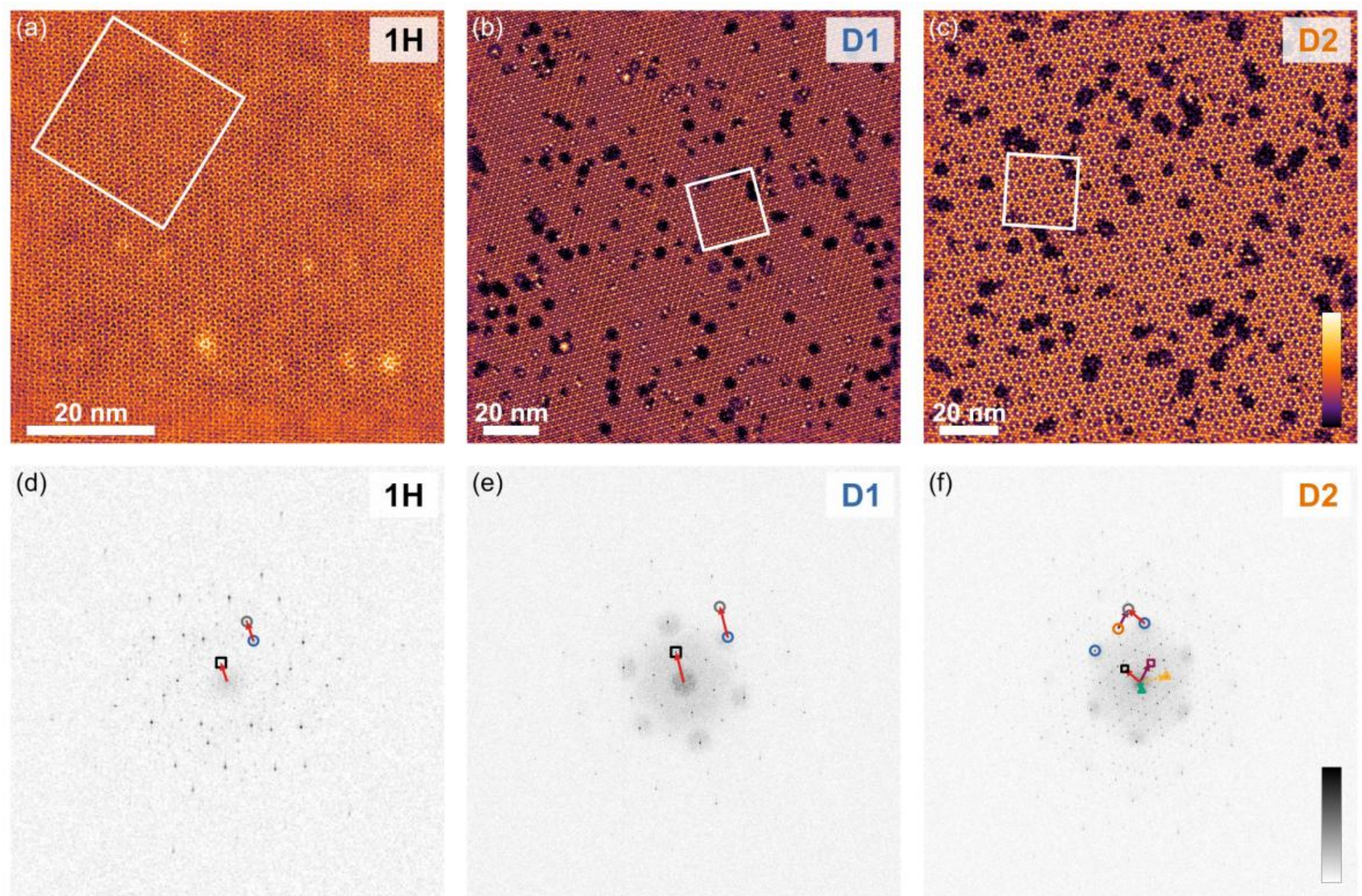

Fig. S3 – **Large area differential conductance images.** (a-c) show the full images of the differential conductance from where Fig.2(a-c) were extracted. The white squares indicate the areas shown in Fig.2(a-c). (d-f) Fourier transform of (a-c), respectively, only with windowing applied. (Map parameters: (a) $f_{drv}$ = 640 Hz, $V_{drv}$ = 500 mV, $I_{max}$ = 685 pA, $\delta q$ = 0.009 Å$^{-1}$; (b) $f_{drv}$ = 640 Hz, $V_{drv}$ = 500 mV, $I_{max}$ = 1.2 nA, $\delta q$ = 0.004 Å$^{-1}$; (c) $f_{drv}$ = 640 Hz, $V_{drv}$ = 40 mV, $V_b$ = 35 mV, $I_{max}$ = 1.2 nA, $\delta q$ = 0.004 Å$^{-1}$).

The differential conductance images shown in Fig.2 of the main manuscript were cropped from large area maps. The complete maps are shown in Fig. S3(a-c). The respective Fourier transforms are shown in Fig. S3 (d-f), with the relevant peaks highlighted. The images shown in Fig. 2(d-f) of the main manuscript correspond to Fig. S3 (d-f) after 6-fold symmetrization and aligning $\boldsymbol{q}_{0T0}$ or $\boldsymbol{q}_{T00}$ with the y-axis.

The peaks corresponding to the CDW on the 1H and 1T layers can also be observed in topographies. In Fig. S4 we show large area topographies taken on 1H-terminated surface (Fig. S4(a)) and on regions D1 and D2 on a 1T-terminated surface (Fig. S4(b,c)). The respective Fourier transforms shown in Fig. S4(d-f) show the respective CDW peaks and moiré peaks, highlighted using the same notation as introduced in the main manuscript.

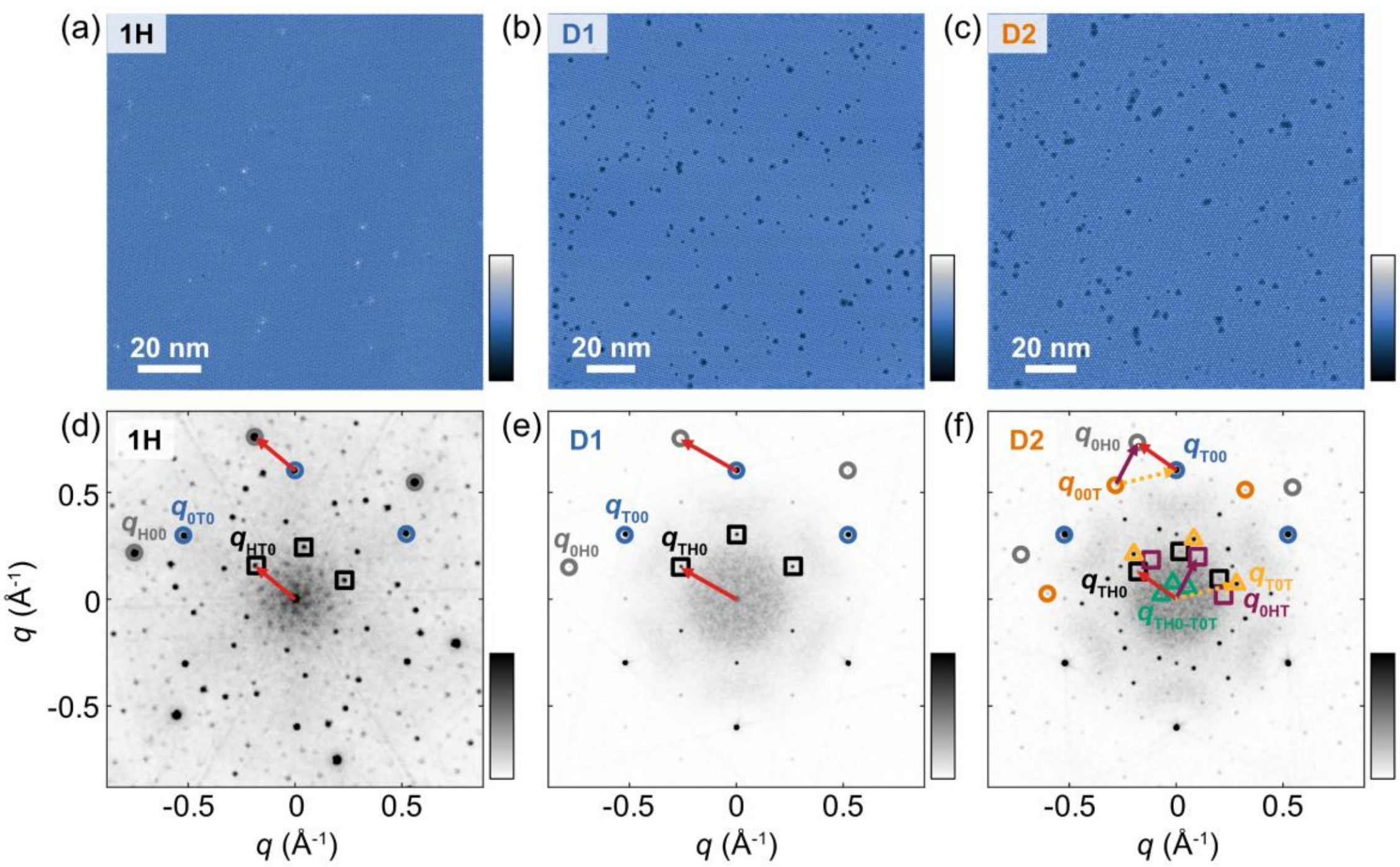

Fig. S4 – **Moiré patterns extracted from topographies**. (a) Large-area topography taken on a 1H-terminated surface, taken in constant-current mode. (b,c) Large-area topographies taken on 1T-terminated surfaces with region D1 (b) and region D2 (c). These were taken in AC-STM mode, with the feedback applied for constant amplitude of the first harmonic at 900mV. (d-f) Fourier transform of (a-c), respectively, after windowing, 6-fold symmetrisation and aligning $\boldsymbol{q}_{0T0}$ or $\boldsymbol{q}_{T00}$ with the y-axis. The same peaks as the ones identified in Fig.2 of the main manuscript are highlighted, using the same notation as in the main manuscript. (Topographies parameters: (a,d) $V_{set}$ = 25 mV, $I_{set}$ = 51 pA, $\delta q$ = 0.005 Å$^{-1}$; (b-f) $f_{drv}$ = 640 Hz, $V_{drv}$ = 40 mV, $V_b$ = 35 mV, $I_{max}$ = 1.2 nA, $\delta q$ = 0.004 Å$^{-1}$).

### S5. Simulation of super modulations

To determine what combination of periodicities can give rise to the peaks found on the 1H-surface and 1T-surface on both regions D1 and D2, we have simulated the moiré pattern between the different periodicities found in this material. To do that, we used the formalism introduced in ref. (*10*):

$$f_i(\boldsymbol{r}) = \frac{1}{9} + \frac{8}{9}\cos\left(\frac{1}{2}\boldsymbol{k}_{1,i}\boldsymbol{r}\right)\cos\left(\frac{1}{2}\boldsymbol{k}_{2,i}\boldsymbol{r}\right)\cos\left(\frac{1}{2}\boldsymbol{k}_{3,i}\boldsymbol{r}\right),$$

With the lattice wavevectors given by:

$$\boldsymbol{k}_{1,i} = \frac{2\pi}{a_i}\left[1, -\frac{1}{\sqrt{3}}\right], \qquad \boldsymbol{k}_{2,i} = \frac{4\pi}{a_i\sqrt{3}}[0,1], \qquad \boldsymbol{k}_{3,\mathrm{i}} = \frac{2\pi}{a_i}\left[-1, -\frac{1}{\sqrt{3}}\right],$$

Where $a_m$ is the lattice constant of the periodicity we are using. To calculate the moiré pattern we calculate

$$f_{\text{moiré}}(\boldsymbol{r}) = f_1^{n1}(\boldsymbol{r}) \times f_2^{n2}(\boldsymbol{r}) \times \cdots \times f_i^{nj}(r),$$

Where $n1, n2, \ldots, nj = 1,2, \ldots$ are indices to account for higher order terms up to the j$^{\text{th}}$ order, and *i* is the number of lattices over which to calculate the moiré pattern. In Fig.2 of the main manuscript, we calculate $f_{\text{moiré}}(\boldsymbol{r})$ for up to three lattices. For the images taken on the 1H-termination (Fig.2(d) and Fig.S5), we consider the atomic lattice and 1H-CDW of the surface layer, and 1T-CDW of the subsurface layer, while for the images taken on the 1T-termination we account for the CDWs on each layer of the three-layer stacks of 1T-1H-1T (Fig.2(e,f)). To account for angles between the different lattices, we multiply the lattice wave vectors of the respective lattice by the rotation matrix $\boldsymbol{R}(\theta)$, where $\theta$ is relative to the CDW of the topmost 1T-layer (first subsurface layer for the 1H-terminated surface, and surface layer for the 1T-terminated surface).

Table S2 – Wave vectors and angles of the 1H and 1T CDWS. The 2$^{\text{nd}}$ to 4$^{\text{th}}$ columns show the wave vectors and angles extracted from the STM topographies and d*I*/d*V* images taken on a 1H-terminated surface, and regions D1 and D2 on a 1T-teminated surface respectively. The 5$^{\text{th}}$ column has the wave vectors and angles expected in the case of both 1H and 1T CDWS are commensurate with the atomic lattice. The angles between the CDW on the 1H layer and the CDW on the 1T layer are given by $\theta_H$ and $\theta_{00T}$, which are relative to $\boldsymbol{q}_{0T0}$ for 1H-1T-1H and relative to $\boldsymbol{q}_{T00}$ for regions D1 and D2. The quantity δq is the q-resolution of the images used to extract the q-values, and $s_H$=100 x ($q_{H00}$-$q_{0H0}$)/$q_{H00}$ that characterizes the contraction/expansion of the CDW on the 1H layer in between 1T layers compared to the CDW q-vectors of the 1H-terminated surface.

| type of surface | 1H-1T-1H | 1T-1H-1T (D1) | 1T-1H-1T (D2) | Ideal Commensurate CDWs |
|---|---|---|---|---|
| $\boldsymbol{q}_H$ [Å$^{-1}$] | 0.779 | 0.797 | 0.754 | 0.726 |
| $\boldsymbol{q}_T$ [Å$^{-1}$] | 0.600 | 0.603 | 0.603 | 0.604 |
| $\boldsymbol{q}_{HT}$ [Å$^{-1}$] | 0.247 | 0.301 | 0.219 | 0.204 |
| $\|\boldsymbol{q}_H\|/\|\boldsymbol{q}_T\|$ | 1.297 | 1.323 | 1.251 | 1.202 |
| $\|\boldsymbol{q}_{at}\|/\|\boldsymbol{q}_H\|$ | 2.779 | 2.726 | 2.882 | 3.000 |
| $\theta_H$ [°] | 13.90 | 19.17 | 13.90 | 13.90 |
| $\theta_{T00}$ [°] | 0.00 | 0.00 | 0.00 | 0.00 |
| $\theta_{00T}$ [°] | 0.00 | 0.00 | 27.80 | 0.00 |
| $\delta q$ [Å$^{-1}$] | 0.009 | 0.004 | 0.004 | - |
| $s_H$ [%] | 0.0 | -2.3±0.5 | 3.2±0.5 | 6.8 |

In Table S2, we summarize the parameters extracted from the images shown in section S4, which are the parameters used for the simulations shown in Fig.2 of the main manuscript. For comparison, in Table S2 we include the wave-vectors for the case where both the 1T and 1H CDWs are commensurate with the atomic lattice.

To extract the peak positions from the Fourier transform of each image, we fitted a Gaussian function to an area around each characteristic q-vector. To identify the contribution from higher order moiré effects, we show in Fig. S5(a) the Fourier transform of a topography taken at the 1H-terminated surface where the peak positions for the atomic lattice (yellow), 1H-CDW (bue) and 1T-CDW (red) are shown, as well as the first order moiré peaks. In this case, the moiré peaks are evaluated in reciprocal space using directly the extracted q-vectors marked in the image for direct comparison. One can see that not all the peaks that appear in the Fourier transform are captured. Most of the remaining peaks can be accounted for by considering higher order moiré contributions, as shown in Fig. S5(b), where contributions up to 3$^{rd}$ order were considered.

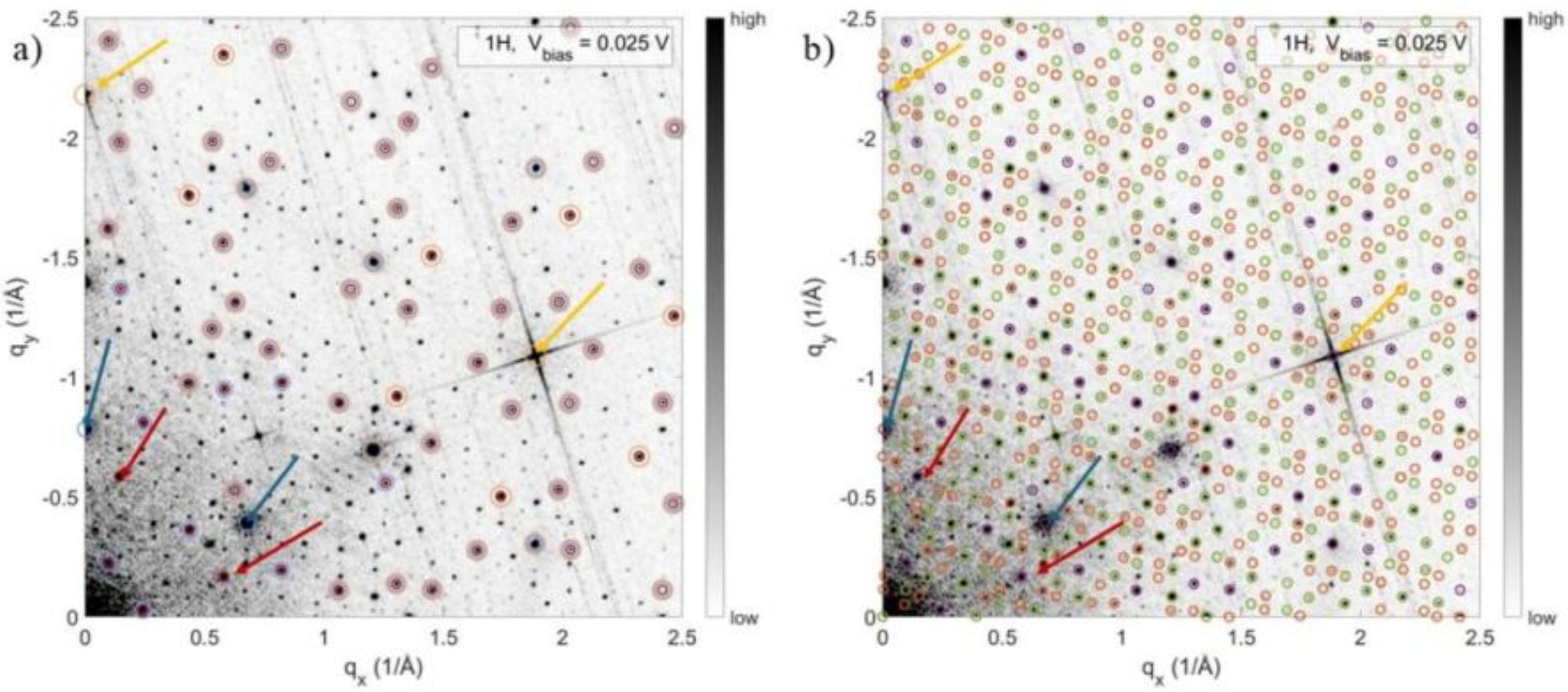

Fig. S5 – **Contribution from higher order moiré effects.** (a) Fourier transform of a topography taken on a 1H-terminated surface at 25mV, with the first order moiré peaks between the atomic, 1H-CDW and 1T-CDW lattices superimposed. The atomic peaks are indicated with a yellow arrow, the 1H-CDW peaks with a blue arrow and the 1T-CDW with a red arrow. The first order moiré peaks are marked with circles coloured according to their components: atomic (yellow), 1H CDW (blue), 1T CDW (red). (b) Same Fourier transform image but now with the first order moiré peaks marked with purple circles, the additional second order moiré peaks marked with green circles and third order moiré peaks are marked with orange circles. There are many distinct moiré peaks up to third order. The Fourier transform image was taken using windowing, was corrected for drift using a linear transformation and aligned such that the atomic peak of the surface sulphur lattice is aligned with the y-axis.

## S6. Energy dependence of moiré peaks

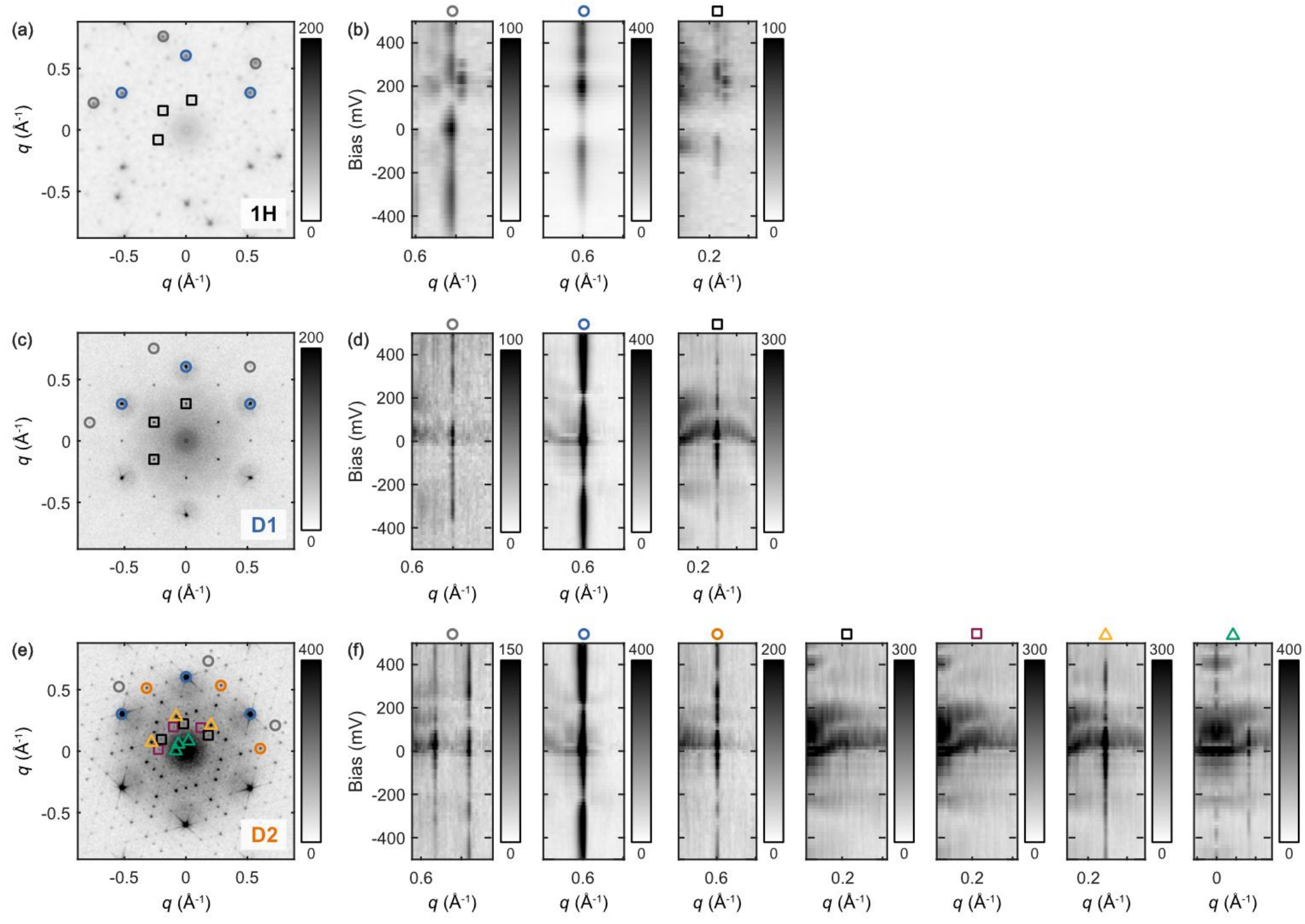

Fig. S6 – **Energy dependence of CDW and moiré peaks.** (a) Fourier transform of a differential conductance map taken at the 1H-terminated surface. (b) energy linecuts across $\boldsymbol{q}_{\mathbf{H00}}$ (gray circle), $\boldsymbol{q}_{\mathbf{0T0}}$ (blue circle) and $\boldsymbol{q}_{\mathrm{HT0}}$ (black square). (c) Fourier transform of a differential conductance map taken at region D1 of 1T-terminated surface. (d) energy linecuts across $\boldsymbol{q}_{\mathbf{0H0}}$ (gray circle), $\boldsymbol{q}_{\mathbf{T00}}$ (blue circle) and $\boldsymbol{q}_{\mathbf{TH0}}$ (black square). (e) Fourier transform of a differential conductance map taken at region D2 of 1T-terminated surface. (f) energy linecuts across $\boldsymbol{q}_{\mathbf{0H0}}$ (gray circle), $\boldsymbol{q}_{\mathbf{T00}}$ (blue circle), $\boldsymbol{q}_{\mathbf{00T}}$ (orange circle), $\boldsymbol{q}_{\mathbf{TH0}}$ and $\boldsymbol{q}_{\mathbf{0HT}}$ (squares), $\boldsymbol{q}_{\mathrm{T0T}}$ (yellow triangle) and $\boldsymbol{q}_{\mathrm{TH0-T0T}}$ (green triangle). (Map parameters: (a,b) $f_{\mathrm{drv}}$ = 640 Hz, $V_{\mathrm{drv}}$ = 500 mV, $I_{\max}$ = 685 pA, $\delta q$ = 0.009 Å$^{-1}$; (b-f) $f_{\mathrm{drv}}$ = 640 Hz, $V_{\mathrm{drv}}$ = 500 mV, $I_{\max}$ = 1.2 nA, $\delta q$ = 0.004 Å$^{-1}$).

## S7. Comparison of conventional QPI and fast-QPI

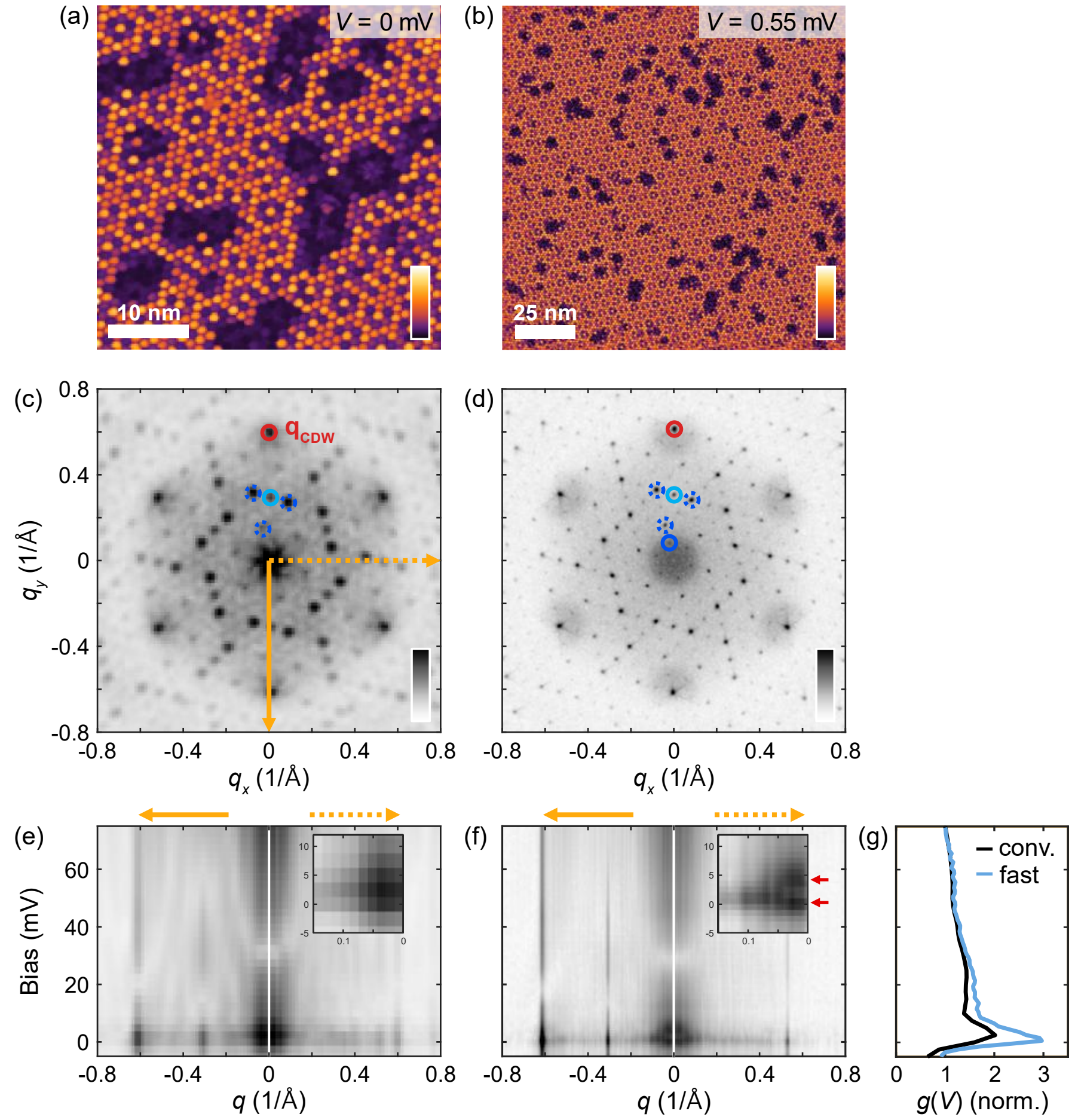

Fig. S7 – **Comparison of QPI map taken conventionally and taken with the fast-approach.** (a) Differential conductance real space image at 0 mV. (b) Real space differential conductance layer taken with the fast-approach at 0.55 mV. (c,d) Fourier transform of (a) and (b), respectively, after correcting for tip distortion and symmetrisation by 6-fold rotation to increase signal to noise ratio. The same peaks are observed in both images. In (d), due to the increased q-resolution, an additional set of peaks can be resolved (full dark blue circle), that cannot be resolved in (c). (e,f) Energy dispersions along the directions indicated by the yellow arrows in (c) for both the conventional (e) and fast (f) QPI maps. It can be seen that the fast-approach reproduces the features observed in the conventional map, but with increased energy and q-resolution as expected. The insets show a zoom in around q=0 1/Å and close to the Fermi level. While in the conventional map only a 'blob' is observed, in the increased resolution obtained due to the fast approach, two distinct dispersions can be resolved. (g) Average differential conductance for both the conventional (black) and fast (blue) QPI maps. It shows that they are consistent with each other.

In this work we used a combination of sparse sampling and fast spectroscopy mapping(*11–13*) to perform the large area QPI maps shown in Fig. S3, and Figs.2,3 of the main manuscript, which we call fast-QPI. To illustrate the improved energy and q-resolution, we compare two maps taken with the conventional mapping and fast-QPI in Fig. S7 taken on the same surface. It shows consistent results, with the Fourier transform showing the same structure, and the spectra looking the same. The energy linecuts highlight the increased energy and q-resolution that allow to resolve the QPI features corresponding to flat bands around $q$=0.